\documentclass[letter]{aa}  

\usepackage{graphicx}
\usepackage{txfonts}
\usepackage{bbold}
\usepackage{gensymb}
\usepackage{amsmath}
\usepackage{stmaryrd}
\usepackage{multirow}
\usepackage[colorlinks=true, citecolor=blue, linkcolor=blue, urlcolor=blue]{hyperref}
\usepackage{fontawesome5}

\begin{document} 

   \title{Forecasting the occupancy of satellite megaconstellations in SKA observations}
   \titlerunning{Satellite occupancy in SKA observations}

   \author{N. Cerardi\inst{1}\fnmsep\thanks{\email{nicolas.cerardi@epfl.ch}}
          \and
          E. Tolley\inst{1}
          \and 
          F. di Vruno \inst{2}
          }
    \authorrunning{Cerardi, N., et al.}

   \institute{Institute of Physics, Laboratory of Astrophysics, École Polytechnique Fédérale de Lausanne (EPFL), Switzerland
         \and Square Kilometre Array Observatory, Lower Withington, Macclesfield, Cheshire, SK11 9FT, UK
             }

   \date{Received 28/11/2025; accepted 20/04/2026}

  \abstract 
   {The Square Kilometre Array (SKA) is expected to start science operations in 2030 and by that time there could be up to 10$^5$ artificial satellites in Earth's orbit, comprising an increase of an order of magnitude compared to 2024. Most of these new satellites will belong to satellite megaconstellations aimed at providing communication services all over Earth. These satellites create radio frequency interference (RFI) that can impact the observations of modern radio telescopes. In this Letter, we forecast the amount of observing time for which the SKA interferometers will be exposed to satellites, risking RFI contamination. We employed an analytical model and considered two cases of exposure to satellites; (1) satellites that only lie in the main beam and (2) satellites that lie in the main beam or the first sidelobe. We show that for SKA-Low, the exposure is high, with satellites in the beam for 30\% of the observation time across half of the frequency range, rising up to 100\% below 100 MHz. For SKA-Mid, high frequencies are mostly spared, but observations below 1 GHz could also end up seeing satellites for at least 30\% of the time. We conclude that satellites will be unavoidable during SKA observing conditions, risking a strong impact on the RFI environment. This will necessitate a concerted effort to obtain accurate measurements of satellite RFI and to improve our understanding of the impact on various science cases. Finally, new mitigation techniques that are less data-destructive than simple flagging must be introduced.}


   \maketitle


\section{Introduction}

Over the next decade, the Square Kilometre Array (SKA) is set to revolutionize radio astronomy thanks to its unprecedented sensitivity. The SKA is comprised of two interferometers, SKA-Low, built in Australia and covering the 50-350 MHz range, and SKA-Mid, built in South Africa and covering the 0.35-15.4 GHz range. SKA promises to answer key questions in astrophysics and cosmology, including the epoch of reionization \citep{Koopmans_2015}, gravitational waves \citep{janssen2014gravitationalwaveastronomyska}, star formation \citep{Hoare:2015Oh}, and galaxy evolution, and other key topics in the field. 

\par Radio telescopes such as the SKA are subject to radio frequency interference (RFI) that corrupt astrophysical signals. Common mitigation measures include building such facilities in extremely remote locations and flagging remaining RFI. However, RFI from artificial satellites represent a growing concern for the community \citep[e.g.][]{Grigg_2025}, as the number of objects annually launched into orbit has increased by a factor of 5 since 2020\footnote{\url{www.esa.int/Space_Safety/Space_Debris/ESA_Space_Environment_Report_2024}}. These objects are primarily small communications satellites that make up the megaconstellations deployed by private operators. 
Satellite RFI can have a number of possible origins: (i) intended transmission in the bands allocated by the Radiocommunication sector of the International Telecommunication Union (ITU-R); (ii) unintended electromagnetic radiation (UEMR), from the spacecraft electronics; and (iii) reflection from bright terrestrial sources. All three interference mechanisms have been detected by various radio telescopes around the world \citep{Di_Vruno_2023, Grigg_2023, bassa2024brightunintendedelectromagneticradiation, Zhang_2025}. In particular, UEMR is especially problematic, as it combines both strong spectral lines and broad band features across the 20-200 MHz regime, a band of frequencies that will be crucial for studies of the cosmic dawn \citep{Bera_2023}.

\par While the astronomical community is actively collaborating with the satellite industry to find mitigation measures, particularly with respect to the problem of UEMR and intentional transmissions, it is necessary to estimate the amount of contamination for different mitigation scenarios. Direct simulations are challenging because they require propagating around 50000 satellites on their orbits with a sub-second resolution, as satellites orbiting at $\sim$500km or less typically cross the field of view (FoV) in a few seconds at $\sim$200MHz. The emitters' radio beams and directionality must also be simulated, which adds yet another layer of complexity.  Instead, one can estimate the average density of satellites in each direction in the sky for a given observatory. \cite{Bassa_2022} developed such a method and predicted the number of satellite trails in optical telescopes. However, no such prediction exists for radio telescopes, despite the importance of understanding and modeling the future conditions of observatories such as the SKA.

\par In this Letter, we forecast the rate of satellite encounters in SKA-Low and SKA-Mid observations. We note that our aim here is not to predict the actual power received from satellites, which would determine whether a set of observation times and frequencies would have to be discarded. In Section \ref{sec:model}, we adapt the framework of \cite{Bassa_2022} to model the fraction of time with at least one satellite present in an effective beam. In Section \ref{sec:forecasts}, we present forecasts for both SKA interferometers.

\section{Model}\label{sec:model}

\subsection{Satellite constellations}
Satellite in megaconstellations are launched following the Walker pattern \citep{Walker_1984}. They populate orbital planes that are characterized by their inclination, $i$, altitude, $h$, longitude of their ascending node, $\Omega$, and their number of satellites. Regrouping all planes with same $i$ and $h$ parameters constitutes an orbital shell, populated by a total of $n$ satellites. In this work, we take into account the large fleets filed to ITU-R: Starlinks Phase 1 (4408 satellites) and Phase 2 (29988), Oneweb (720), GuoWang (12992), QianFan (1296), and Leo (formerly Kuiper, 3236). The deployment of these megaconstellations has already begun, with the most advanced one being Starlink, with more than 9000 active satellites (February 2026). We detail in Table \ref{tab:shells} the shells composing each megaconstellation and their characteristics $(i,h,n)$, which serve as the input to our statistical model.

\subsection{Satellite counts in the FoV}

The first step of our analytical model is to compute the expected number of satellite counts in the FoV for a specific observation. For one single shell, the density of satellites in an element of solid angle, d$\Omega$, and observed from a latitude, $\phi_{\rm o} $, is given by \citep{Bassa_2022} and expressed as
\begin{equation}\label{eq:rhosat}
    \rho_{\rm sat} = n ~ P(\phi_{\rm s}, i, h)~  \frac{d^2 \rm{d} \Omega}{\cos{\theta}}, 
\end{equation} 
with $\theta$ as the impact angle, $d$ the distance to the observed segment of orbital shell, $\phi_{\rm s}$ its latitude, and $P(\phi_{\rm s}, i, h)$ the probability density for a single satellite in the shell to be at the latitude, $\phi$. Then, the expected number of observed satellites in the shell $N_{\rm shell}^{\rm obs}$ can be expressed as
\begin{equation}\label{eq:nobs}
    N_{\rm shell}^{\rm obs}=\rho_{\rm sat}~ \biggl(\frac{\pi}{4} ~ L_{\rm FoV}^2+\omega_{\rm sat} ~L_{\rm FoV}~t_{\rm obs}\biggr),
\end{equation}
with $\omega_{\rm sat}$ the average angular velocity of satellites, $L_{\rm FoV}$ the diameter of the instrument FoV, and $t_{\rm obs}$ the duration of the observation. Equation \ref{eq:nobs} assumes that satellites follow straight lines across the FoV; to check for the validity of this assumption, we computed realistic satellite trajectories and found that this is still reasonable for the largest main beam size considered here. The formulas to determine $P(\phi_{\rm s}, i, h)$, $\omega_{\rm sat}$, $\theta$, and $d$ are given in Sect. \ref{app:analyticalmodel}.

\begin{figure}
    \centering
    \includegraphics[width=0.98\linewidth, trim=.6cm 0 1.6cm 0.4cm, clip]{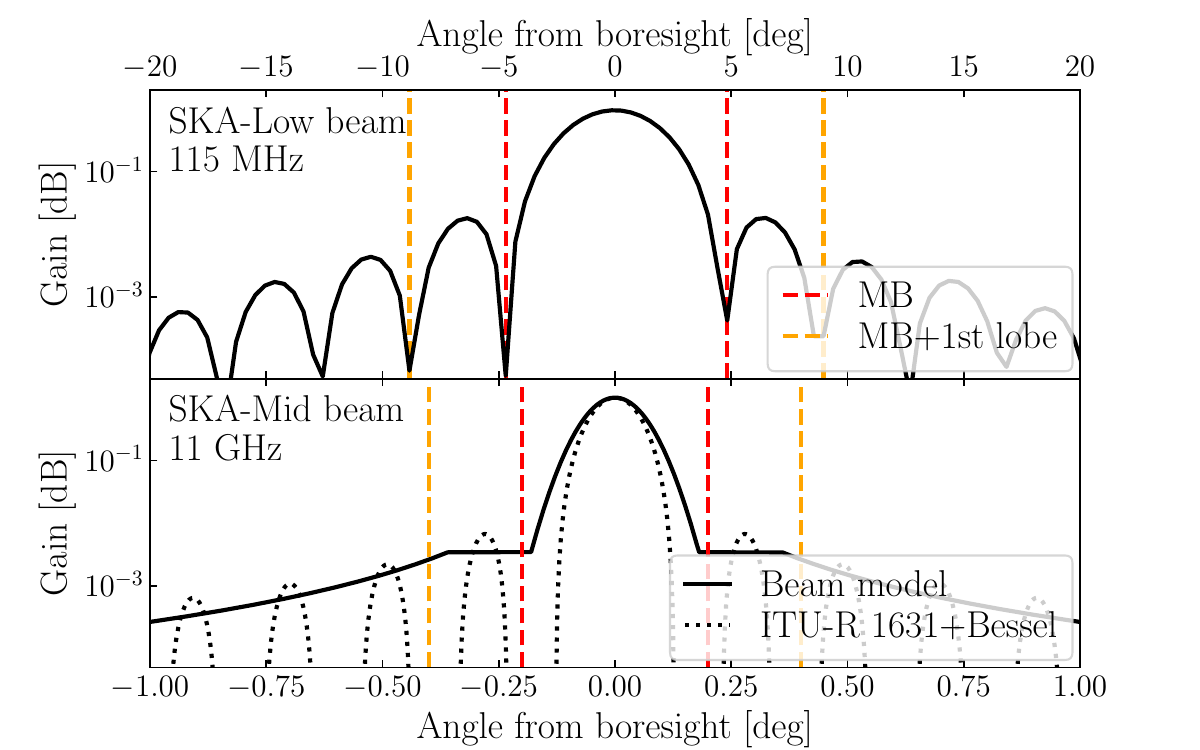}
    \caption{Beam models (solid black) adopted for SKA-Low (top) and SKA-Mid (bottom). The effective beams defined until the first (red) and second (orange) null are also shown, where MB stands for main beam.}
    \label{fig:beammodels}
\end{figure}

\par To determine $L_{\rm FoV}$,  we used a primary beam radial model and definde a circular effective beam. In practice, the exact value of $L_{\rm FoV}$ depends on the RFI brightness and the required sensitivity of the observation. In this Letter, we determined a forecast for two independent scenarios: one where the effective beam consists of the main beam only (e.g., until the first null) and one including the first lobe (e.g., until the second null). 
For SKA-Low, we used the beam model for a single station from \cite{bonaldi2025squarekilometrearrayscience}, whereas for SKA-Mid, we used the parabolic single-dish model from ITU-R recommendation RA.1631\footnote{\href{https://www.itu.int/rec/R-REC-RA.1631/en}{www.itu.int/rec/R-REC-RA.1631/en}}. Hence, for both models, we estimated the radii of the first and second nulls and obtained $L_{\rm FoV}$, as shown in Figure \ref{fig:beammodels}, for specific frequencies. At each frequency, we rescaled the effective beamwidth with $L_{\rm FoV} \propto \nu^{-1}$.

\par
\begin{figure}
    \centering
    \includegraphics[width=0.7\linewidth, trim=0 .3cm 0 .2cm, clip]{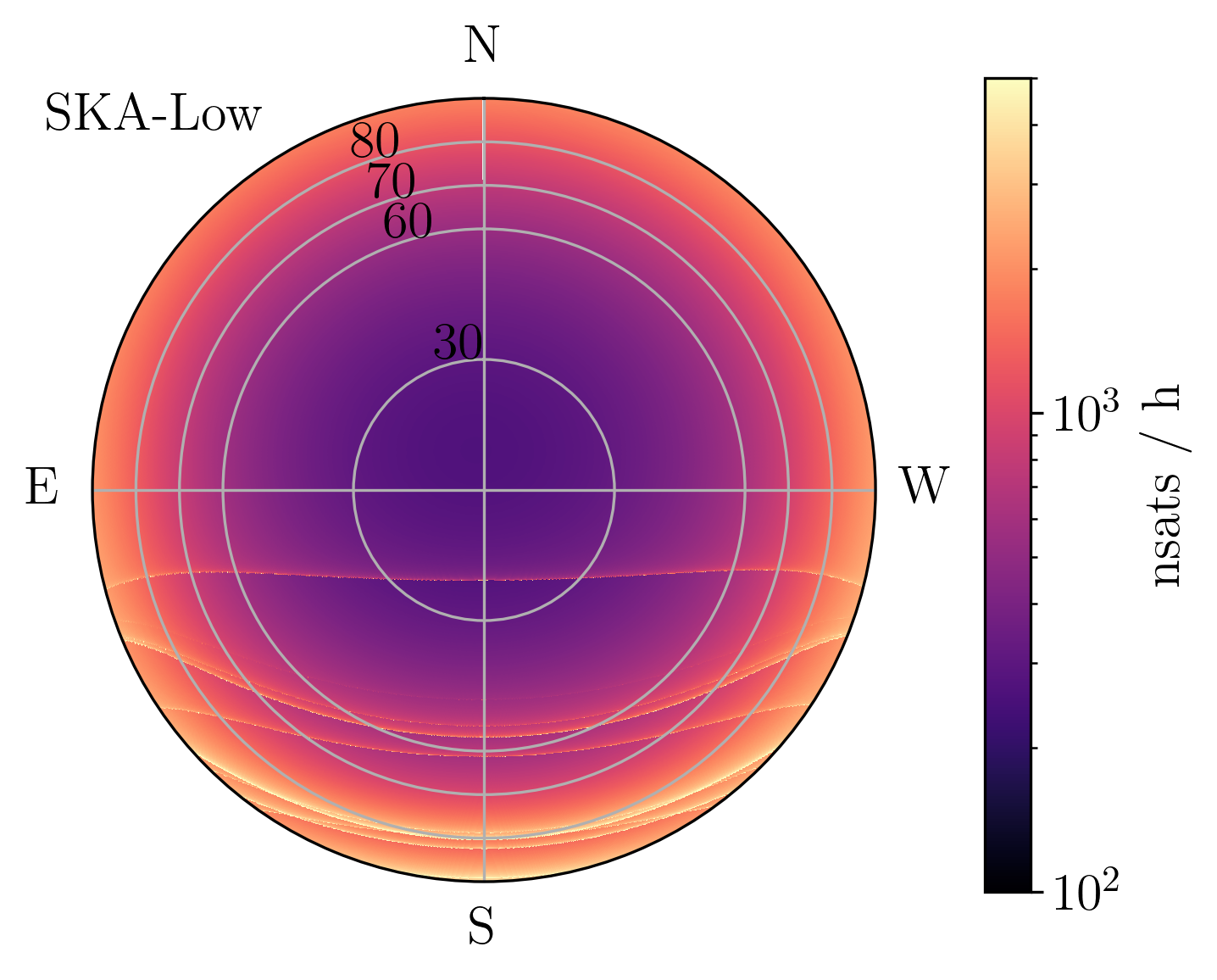}
    \includegraphics[width=0.7\linewidth, trim=0 .33cm 0 .2cm, clip]{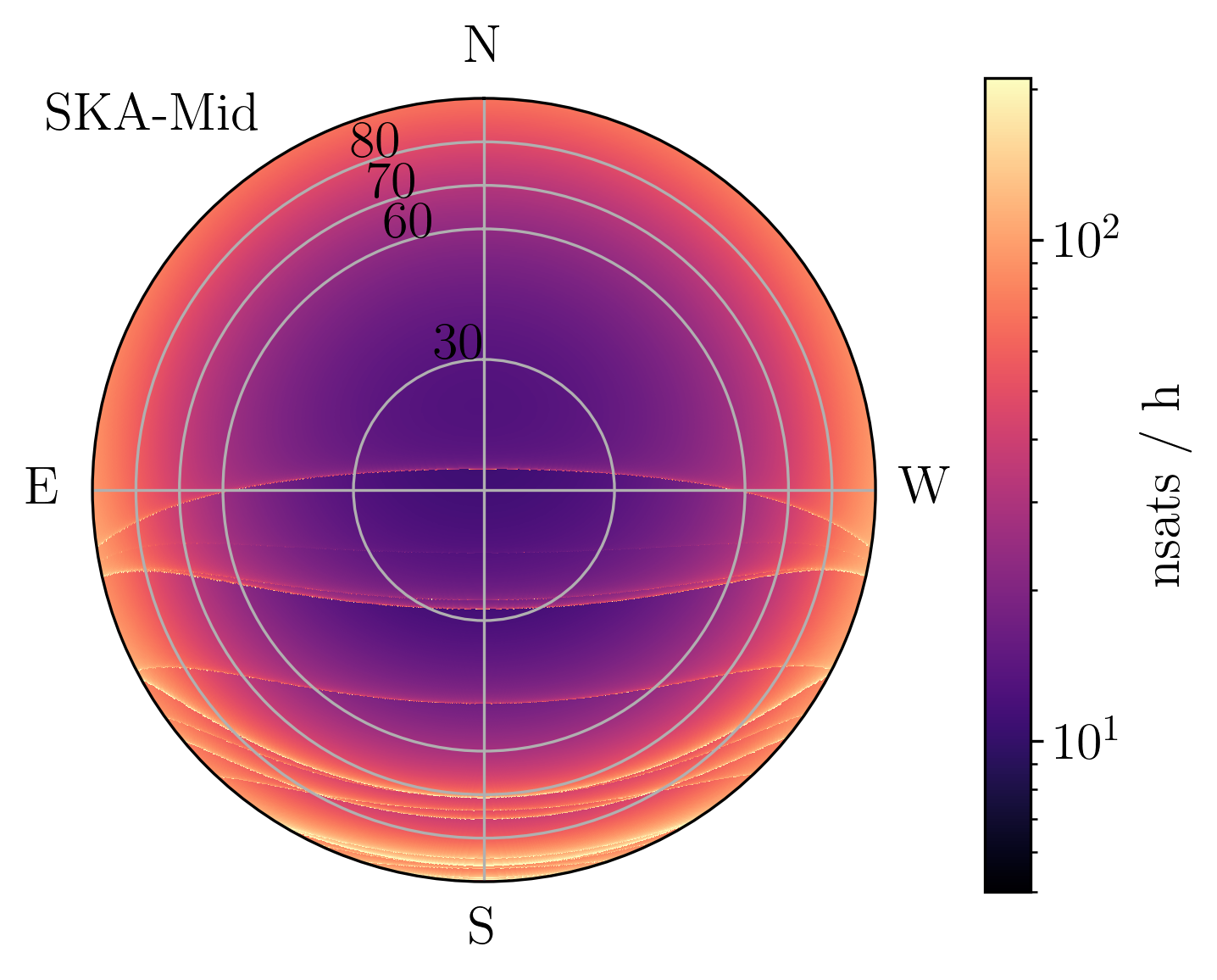}
    \caption{Number of satellites per hour as a function of the pointing direction, for the SKA-Low (top) and SKA-Mid (bottom) sites, respectively, with $L_{\rm FoV}=9.5\degree$ ($\sim115$ MHz) and $0.4\degree$ ($\sim11$ GHz). Note: the colorbar differs between the two maps.}
    \label{fig:skymaps}
\end{figure}

 Equation \ref{eq:nobs} could then be evaluated for each input shell and summed to obtain a map of the total number of satellites expected, $N_{\rm tot}^{\rm obs}$, as a function of pointing direction for a particular observation. In Fig. \ref{fig:skymaps} we show these maps for SKA-Low (top) and SKA-Mid (bottom), with $t_{\rm obs}=1$ h. With their respective latitudes of $26\degree$S and $30\degree$S, the interferometer sites are below many of the orbital shells defined in Table \ref{tab:shells}. High-density lines in the maps originate from the edges of the shells with low $i$. We note that the results scale with $L_{\rm FoV}$, which we considered here until the first null, e.g. $9.5 \degree$ for SKA-Low (beam size at $\sim115$ MHz) and $0.4\degree$ for SKA-Mid (beam size at $\sim11$ GHz). This explains the different orders of magnitudes seen among the two panels. The main takeaway from these maps is that the average exposition to satellites do not appear uniform on the sky and this can impact some astronomical targets more than others due to the orbital structure of each megaconstellation.

Using the main beam as the effective beam may underestimate $L_{\rm FoV}$ based, for instance, on the example of Starlink satellites, which often show broadband radiation on the order of 1 Jy \citep{Di_Vruno_2023}, and the brightest satellites are detected and flagged even when they are located in the sidelobes. We also note that RFI contamination can still occur outside of these effective beams: the second and third sidelobes exceed -30 dB and could pick up bright RFI sources. Faint RFI sources might be individually undetectable at present, but could have significant cumulative effect in the future.

\subsection{Fraction of observing time with at least one satellite}

The last step of the analytical model is to generate mock catalogues of satellite crossings. For the $k$-th shell, we sample $X_k \sim {\rm Poisson}(N_{\rm shell}^{\rm obs})$ and generate a list of random ingress times, $t_{k,j}$, and crossing durations, $\Delta t_{k,j}$, for $j \in \llbracket 1, X_k \rrbracket$ (see Sect. \ref{app:modelcrossings} for more details). Both $X_k$ and $\Delta t_{k,j}$ depend on the model parameters $L_{\rm FoV}, i, n$, and $h$. From these quantities, we can derive $N_{\rm sat}(t)$, the instantaneous number of satellite in the effective beam, 
\begin{equation}\label{eq:nsat_t}
    N_{\rm sat}(t) = \sum_{k ~ {\rm shells}} \sum_{j=1}^{X_k}\mathbb{1}_{[t_{k,j}, t_{k,j}+\Delta t_{k,j}[} \left( t \right),
\end{equation}
with $\mathbb{1}$ representing the indicator function. Finally, we estimate $f_{N_{\rm sat}\geq1}$, which is the fraction of observing time with at least one satellite in the effective beam. For statistical robustness, we bootstraped our model 100 times and calculate the median. In Sect. \ref{app:modelvalidation}, we describe the validation of our analytical approach against discrete orbit simulations. Although the Poisson and uniform sampling of $X_k$ and $t_{k,j}$ might seem too simplistic for well-coordinated fleets, our tests show that our model correctly reproduces the occupancy of satellites in current observing conditions.

 \par Our prediction of $f_{N_{\rm sat}\geq1}$ can be seen as an upper bound on the exposure to RFI, for a single frequency channel. This upper bound could be reached if the two following conditions are filled: (i) if all satellites are RFI emitters on the given channel \citep[in][for instance, broad band RFI was detected for about 40 out of 68 Starlinks, at similar frequencies]{Di_Vruno_2023}; and (ii) if the radiation pattern is isotropic (this could be reasonably valid for UEMR, but not for intended transmissions).

\section{Forecasts}\label{sec:forecasts}

We derived forecasts for the SKA-Low and SKA-Mid telescopes. We recall that results scale with $L_{\rm FoV}$, and we present results across the complete frequency range of each interferometer below for the two scenarios with respect to the effective beam.

\subsection{SKA-Low}

\begin{figure}
    \centering
    \includegraphics[trim=0 .26cm 0 .41cm, width=0.8\linewidth, clip]{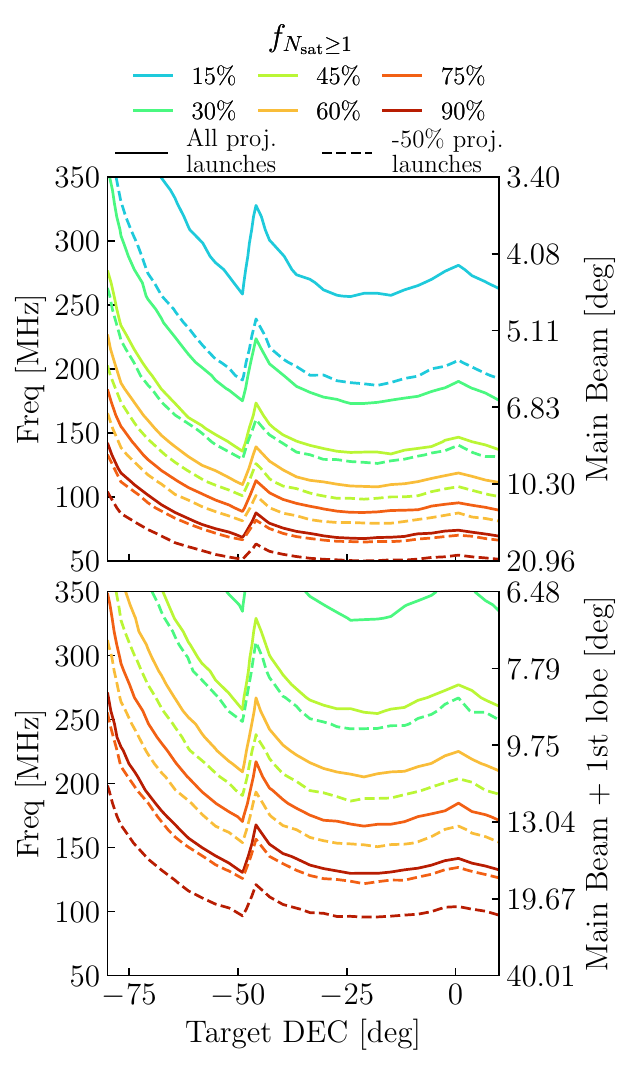}
    \caption{Forecast of the fraction of time exposed to satellites ($f_{N_{\rm sat}\geq1}$) in SKA-Low, as a function of the observing frequency and the declination of the target. It includes megaconstellations from Table \ref{tab:shells}. The top panel shows the analysis counting satellites only in the main beam, whose size varies with frequency. The bottom panel shows the analysis including satellites also in the first sidelobe. Solid lines show the contour for the analysis with all satellites, while dashed lines indicates the contours for the analysis with -50\% projected launches.}
    \label{fig:maps_skalow}
\end{figure}

We first present results for the SKA-Low telescope. At 115 MHz, we calculate $L_{\rm FoV}=9.5\degree$ until the first null and $L_{\rm FoV}=17.8\degree$ until the second null. In Figure \ref{fig:maps_skalow}, we show the forecasts of $f_{N_{\rm sat}\geq1}$ for SKA-Low, for the main beam only (top) and when including the first sidelobe (bottom), as a function of the beam size (also including the corresponding frequency) and the declination of the observation. We show the 15, 30, 45, 60, 75, and 90\% isocontours as solid lines. As expected we observe increased occupancy at low frequencies as the beam gets wider. In addition, $f_{N_{\rm sat}\geq1}$ also increases as the observation pointing moves toward the south celestial pole, as observations get closer to the dense Starlink shell at $i=43\degree$. The peak at $-45\degree$ is induced by the future GuoWang shell at $i=30\degree$. We highlight that all SKA-Low observations will be highly exposed, with a satellite present in the main beam or the first sidelobe at least 30\% of the time, and up to 90\% below 150 MHz. However, these results are highly dependent on the population of LEO satellites. If not all satellites listed in the ITU filings are launched the situation would improve. Therefore, we ran another analysis reducing the expected population of satellites by 50\% for all fleets, except for ones that are already complete (e.g. Starlink Phase 1 and OneWeb) and show the corresponding isocontours in dashed lines. This amounts to 28884 satellites, compared to the total of 52640 given in the initial analysis.
Under this assumption, $f_{N_{\rm sat}\geq1}$ decreases by at most 25\%, depending on the frequency and target DEC. Satellites would be in the main beam (resp. and 1st sidelobe) at least 45\% of the observation time under 100 MHz (200MHz). From both analyses, we conclude that discarding all observed data whenever a satellite is in the main beam would be a very detrimental strategy for SKA-Low.

\subsection{SKA-Mid}

\begin{figure}
    \centering
    \includegraphics[trim=0 .26cm 0 .41cm, width=0.8\linewidth, clip]{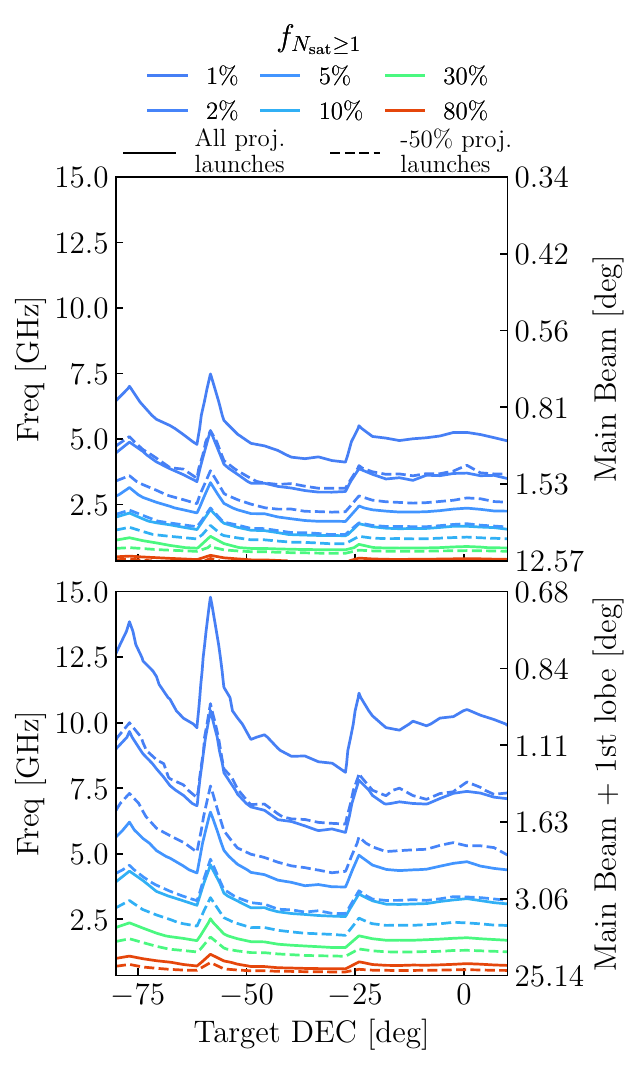}
    \caption{Forecast of the fraction of time exposed to satellites, ($f_{N_{\rm sat}\geq1}$), in SKA-Mid, as a function of the observing frequency and the declination of the target. It includes megaconstellations from Table \ref{tab:shells}. The top panel shows the analysis counting satellites in the main beam only, which size varies with frequency. The bottom panel shows the analysis including satellites also in the first sidelobe.}
    \label{fig:maps_skamid}
\end{figure}

We conducted the same analysis for SKA-Mid, with $L_{\rm FoV}=0.4\degree$ when we were considering only the main beam and $L_{\rm FoV}=0.8\degree$ when including the first sidelobe, at 11 GHz. Figure \ref{fig:maps_skamid} shows the counterpart of Figure \ref{fig:maps_skalow} for SKA-Mid. As SKA-Mid spans two orders of magnitude in frequency, there are important variations in $f_{N_{\rm sat}\geq1}$, following the same trends that for SKA-Low. While high frequencies have satellites in the telescope’s FoV for a low percentage of time, at low frequencies, $f_{N_{\rm sat}\geq1}$ can potentially reach 100\%. Below 2.5 GHz, including the first sidelobe in the analysis, 20\% of the observations could be impacted, up to 80\% below 500 MHz. The stripe due to the $i=30\degree$ GuoWang shell is still present, but shifted to higher declination, as SKA-Mid is further south than SKA-Low. As before, we carried out a second analysis by reducing the amount of satellites by -50\% for the incomplete fleets (see dashed isocontours in the figure). Even with a 50\% cut in the launches to come, up to 30\% of the observing time could have a satellite in the main beam and 1st sidelobe, under 1 GHz. From Figures \ref{fig:maps_skalow} and \ref{fig:maps_skamid}, we highlight that 21cm science could be severely affected, for all redshifts of interest. 

\section{Conclusions}
\par We present our forecasts for the fraction of time when the SKA-Low and SKA-Mid will encounter a satellite within their effective FoV. We used an analytical method that allowed us to avoid the use of orbit simulations and adapted it in the context of radio telescopes. Our forecasts are based on existing megaconstellations and expected launches. We found that SKA-Low would be strongly exposed to these satellite fleets, with $f_{N_{\rm sat}\geq1}$ results ranging from 10\% (at 300MHz) to 80\% (100MHz). These numbers increase, respectively, to 30\% and 100\% when including the first sidelobe in the analysis. We find the satellite occupancy in SKA-Mid to be below 5\% above 5 GHz under our assumptions, but at least 30\% below 1 GHz. Our code is publicly available\footnote{    
\href{https://github.com/nicolas-cerardi/Analytical-Satsky}{ github.com/nicolas-cerardi/Analytical-Satsky \faGithubSquare}  } and can be easily reused for other telescopes. The use of our tool to derive statistics for values beyond $f_{N_{\rm sat}\geq1}$ should be straightforward.
\par A great deal of uncertainty remains concerning the level of RFI coming from megaconstellations, mostly due to poor knowledge of UEMR (e.g., its dependence on the satellite orientation or on its instantaneous activity). Robust measurements in greater numbers are necessary, as well as precise simulations of interferometric observations with RFI to improve our understanding of its exact impact on different science cases. For instance, it is important to investigate whether the aggregate RFI signal from hundreds of satellite completely incoherent and only ends up increasing the level of noise or whether these satellites will contribute to a coherent signal. In addition, we must consider the impact of satellites following specific trajectories on observations, particularly in the context of power spectrum measurements. We plan to study these key questions in future works.

\par The deployment of satellite megaconstellations might challenge our ability to exploit the full capacities of the SKA.
In any case, strong efforts are needed towards new RFI mitigation techniques, for instance, boresight avoidance \citep{nhan2024spectrumcoexistencedemonstrationeffectiveness} or RFI subtraction \citep[][]{Finlay_2023,finlay2025tabascaliiremovingmultisatellite}. We stress that both methods rely on cooperation with private satellite operators (for steering satellite emission in the first case and sharing orbital information in the second). It is therefore important to maintain constructive relations between all stakeholders involved in the radio spectrum.

\begin{acknowledgements}
NC and ET acknowledge support from the Swiss National Science Foundation under the SNSF Starting Grant ``Deep Waves'' (218396). The authors thank Mark Sargent for useful discussions, Anna Bonaldi for sharing the SKA-Low beam model, and Cees Bassa for comments on the draft manuscript.
\end{acknowledgements}

\bibliographystyle{aa} 
\bibliography{bibliography} 
%
%

\begin{appendix}

\section{Satellite shells considered in this study}\label{app:satshells}

\begin{table}[h!]
    \caption{Megaconstellations considered in this work}
    \centering
        \begin{tabular}{c|c|c|c}
      Constellation & $i$ ($\degree$) & $n$ & $h$ (km) \\ \hline \hline
      \multirow{5}{*}{Starlink Phase 1} & 53 & 1584 & 550 \\
       & 53.2 & 1584  & 540 \\
       & 70 & 720 & 570 \\
       & 97.6 & 348 & 560 \\
       & 97.6 & 172  & 560 \\ \hline
      \multirow{9}{*}{Starlink Phase 2} & 53 & 5280  & 340 \\ 
       & 46 & 5280  & 345 \\ 
       & 38 & 5280  & 350 \\ 
       & 96.6 & 3600  & 360 \\ 
       & 53 & 3360  & 525 \\ 
       & 43 & 3360  & 530 \\ 
       & 33 & 3360  & 535 \\ 
       & 148 & 144  & 604 \\
       & 115.7 & 324  & 614 \\\hline
      OneWeb   & 87.9  & 720  & 1200 \\\hline
      \multirow{7}{*}{GuoWang}  & 85    & 480  & 590 \\
        & 50    & 2000 & 600 \\
        & 55    & 3600 & 508 \\
        & 30    & 1728 & 1145 \\
        & 40    & 1728 & 1145 \\
        & 50    & 1728 & 1145 \\
        & 60    & 1728 & 1145 \\ \hline
      QianFan  & 89    & 1296 & 1160 \\ \hline
      \multirow{3}{*}{Leo}  & 51.9    & 1156 & 630 \\ 
        & 42    & 1296 & 610 \\ 
        & 33    & 784 & 590 \\ 
      \hline \hline
      \multirow{5}{*}{Starlink 2025$^a$} & 43.00 & 2575 & 503 \\
       & 53.15 & 439  & 351 \\
        & 53.15 & 1021 & 466 \\
        & 53.15 & 2383 & 543 \\
        & 70.00 & 432  & 553 \\
        & 97.66 & 231  & 558 \\ \hline
    \end{tabular}\label{tab:shells}
    \tablefoot{$^a$Starlink 2025 refers to the set of Starlinks in orbit in March 2025, according to \textsc{SpaceTrack},binning the individual satellites in seven shells. These shells are only used for the model validation, see Sect. \ref{app:modelvalidation}.}
\end{table}

In this work, we consider five megaconstellations that are currently being deployed and we make use of the orbital configurations described in their respective ITU-R filings. This led to the 26 input shells reported in the top part of Table \ref{tab:shells}. The sources are the following: for Starlink Phase 1, FCC report \href{https://fcc.report/IBFS/SAT-MOD-20200417-00037}{SAT-MOD-20200417-00037}; for Starlink Phase 2, FCC report \href{https://fcc.report/IBFS/SAT-AMD-20210818-00105}{SAT-AMD-20210818-00105}; for OneWeb, FCC report \href{https://fcc.report/IBFS/SAT-LOI-20160428-00041}{SAT-LOI-20160428-00041}; for GuoWang, ITU filing CHN2020-33663; for QianFan, ITU filing \href{https://www.itu.int/ITU-R/space/asreceived/Publication/DisplayPublication/49732}{CHN2023-60476}; for Leo, FCC report \href{https://fcc.report/IBFS/SAT-LOA-20190704-00057}{SAT-LOA-20190704-00057}. The bottom part of the table, corresponding to the so-called Starlink 2025 constellation, summarizes into rough shells the distribution of Starlink satellites in orbit in March 2025. We use these to validate our analytical model against discrete orbital simulations (see Sect. \ref{app:modelvalidation} for more details).

\section{Full description of the analytical model}\label{app:analyticalmodel}  

Here, we give the equations used in our model. Their derivation are detailed in \citep{Bassa_2022}, so we simply report each computation step used in our model, for completeness. We consider an observer located on Earth at longitude $\varphi_{\rm o} $ and latitude $\phi_{\rm o} $, and pointing at sky coordinates ($\alpha$, $\delta$). Without loss of generality, we can assume it lies at the longitude $\varphi_{\rm o}=0 \degree$ (which implies that $\alpha$ is the local hour angle instead of the right ascension).

\subsection{Satellite density in a single shell}\label{sec:sat_dens}

For a shell of satellites at altitude $h$, the line of sight intersects the shell in two points, and their distances to the observer are the roots of the equation,
 \begin{equation}
         d^2 + 2 R_\oplus (\cos \delta ~ \cos \alpha ~ \cos\phi_{\rm o}  + \sin\delta ~ \sin\phi_{\rm o} )d - (h^2 + 2 R_\oplus~  h) = 0.
 \end{equation}
We retain the positive root only (corresponding to the point in front of the telescope) and obtain $d$, the distance from the observer to a shell, along the line of sight.

For a single satellite orbiting Earth with inclination, $i$, and altitude, $h$, its longitude, $\varphi_{\rm s}$, and latitude, $\phi_{\rm s}$, can be obtained as a function of the observer's position and target,
 \begin{align}
     \phi_{\rm s} & = \arcsin \left( \frac{ d \sin\delta +R_\oplus \sin\phi_{\rm o}  }{R_\oplus + h}\right), \\
     \varphi_{\rm s} & =  \text{atan2}\left(\frac{d\cos\delta\sin\alpha}{\cos\phi_{\rm s}  ~ (R_\oplus+h)},
     \frac{d\cos\delta\cos\alpha+R_\oplus\cos\phi_{\rm o} }{\cos\phi_{\rm s}  ~ (R_\oplus+h)} \right).
 \end{align}
 
Then, the probability for it to instantaneously lie at the latitude $\phi_{\rm s}$ is
\begin{equation}
    P(\phi_{\rm s}, i, h) = 
    \begin{cases}
        \frac{1}{2 \pi^2(R_\oplus + h)^2 \sqrt{\sin^2 i - \sin^2\phi_{\rm s}}} & \text{ for } |\phi_{\rm s}|<i,\\
        0 & \text{ otherwise. }
    \end{cases}
\end{equation}
Next, we compute the surface of the orbital shell ($i$, $h$) intersected by an observation of solid angle d$\Omega$. To account for the impact angle $\theta$ (causing the line of sight to not intersect orthogonally the shell), we compute its cosine as
 \begin{equation}
    \cos \theta = \frac{ (R_\oplus + h)^2 + d^2 - R_\oplus^2 }{2d (R_\oplus + h)}.
 \end{equation}
Hence, the surface of the shell intersected by the solid angle d$\Omega$ is $\frac{d^2 \text{d}\Omega}{\cos\theta}$.
We can then multiply it by the number of satellite in the shell and the probability $P(\phi, i, h)$ to obtain equation \ref{eq:rhosat}. 

\subsection{Satellite velocities in a single shell}\label{app:sat_vel}

\par To convert the density of satellites into a number of satellites expected in an observation, we need to take into account the FoV and the motion of satellites. The satellites from one shell and detected in one observation are either orbiting northwards or southwards, with equal probabilities. The longitude of the ascending node in both cases, respectively, is
\begin{align}
    \Omega_{\rm N} & = \varphi_{\rm s} - {\rm arcsin}\left(\frac{\tan\phi_{\rm s}}{\tan i}\right),\\
    \Omega_{\rm S} & = \varphi_{\rm s} + {\rm arcsin}\left(\frac{\tan\phi_{\rm s}}{\tan i}\right) + \pi.
\end{align}
Then, to compute the respective geocentric velocities ${\bf v}_{\rm N}$ and ${\bf v}_{\rm S}$, we build the unit vectors $\bf{CA}$ (from earth center to the ascending node), $\bf CP$ (from earth center to the perigee) and $\bf CS$ (from earth center to the satellite) via
\begin{align}
    {\bf CA} =& \begin{pmatrix}
        \cos\Omega\\ \sin\Omega\\ 0
    \end{pmatrix},  \\
    {\bf CP} =& \begin{pmatrix}
        (\cos\Omega+\frac{\pi}{2})\cos i \\
        (\sin\Omega+\frac{\pi}{2})\cos i \\
        \sin i
    \end{pmatrix}, \\
    {\bf CS} =& \begin{pmatrix}
        \cos\varphi_{\rm s}  \cos \phi_{\rm s}  \\
        \sin\varphi_{\rm s}  \cos \phi_{\rm s}  \\
        \sin \phi_{\rm s}
    \end{pmatrix},
\end{align}
with $\Omega$ being either $\Omega_{\rm N}$ or $\Omega_{\rm S}$. Then, the velocity is computed with:
\begin{equation}\label{eq:v_geo}
    {\bf v} = \sqrt{\frac{G M_\oplus}{R_\oplus+h}} ({\bf CA}\times{\bf CP} )\times{\bf CS}.
\end{equation}
We convert the geocentric velocities into topocentric velocities by subtracting the geocentric velocities of the observer:
\begin{equation}\label{eq:v_topo}
    {\bf v}_{\rm topo} = {\bf v} - \omega_\oplus R_\oplus \cos\phi_{\rm o}  
    \begin{pmatrix}
       -\sin\varphi_{\rm o}  \\ \cos\varphi_{\rm o} \\ 0
    \end{pmatrix},
\end{equation}
with $\omega_\oplus$ the earth angular frequency. Finally, we  project the topocentric velocity orthogonally to the line of sight to obtain the apparent velocity. For this, we use the unit vector from the observer along the line of sight
\begin{equation}
    {\bf OS} = \begin{pmatrix}
        \cos \delta \cos \alpha \\ \cos \delta \sin \alpha \\ \sin \delta
    \end{pmatrix},
\end{equation}
and we obtain
\begin{equation}\label{eq:v_app}
    {\bf v}_{\rm app} = {\bf v}_{\rm topo} - {\bf v}_{\rm topo} \cdot {\bf OS}. 
\end{equation}
In Equations \ref{eq:v_geo}, \ref{eq:v_topo} and \ref{eq:v_app}, all velocities may carry a subscript N or S whether the vectors were built with $\Omega_{\rm N}$ or $\Omega_{\rm S}$. Finally, we compute $\omega_{\rm sat}$ as the average of the apparent velocity norms for the two types of orbit, divided by $d$:
\begin{equation}
    \omega_{\rm sat} = \frac{(||{\bf v}_\text{app, N}||+||{\bf v}_\text{app, S}||)}{2d}.
\end{equation}
In an observation of length, $t_{\rm obs}$, the instantaneous intersection of the FoV with the considered shell is $\pi\left(\frac{L_{\rm FoV}}{2}\right)^2$ and the mean motion of satellites in this shell during the exposure adds a surface of $\omega_{\rm sat}L_{\rm FoV}t_{\rm obs}$, which explains Equation \ref{eq:nobs}.

\subsection{Simulations of satellite crossings}\label{app:modelcrossings}

\par We then construct random realizations of the instantaneous number of satellites present inside the effective beam, $N_{\rm sat}(t)$. We generate mock catalogues of satellites crossings, characterized by the satellites' ingress and flythrough times. The actual number of satellites induced by the shell $k$ can be sampled with $X_k \sim {\rm Poisson}(N_{\rm shell}^{\rm obs})$, which sets the catalogue length for each shell. 
\par Then, the ingress times of the $j$-th satellite from shell $k$, $t_{k,j}$ is simply sampled from the uniform distribution $\mathcal{U}(0,t_{\rm obs})$, with $j \in \llbracket 1, X_k \rrbracket$. 
For $\Delta t_{k,j}$, the duration of the flythrough, we assume that it intersects the FoV with uniformly random direction and ingress time, which leads to the following expression :
\begin{equation}\label{eq:delta_t}
    \Delta t_{k,j} =  \cos \left( ~ \arcsin{\frac{2~Y}{L_{\rm FoV}}} \right) \frac{L_{\rm FoV}}{\omega_{\rm sat}},
\end{equation}
where $Y$ follows the uniform distribution $\mathcal{U}(-L_{\rm FoV}/2, L_{\rm FoV}/2)$. The mock catalogues for all shells were combined to evaluate $N_{\rm sat}(t)$, as expressed in Equation \ref{eq:nsat_t}. Finally, we obtained $f_{N_{\rm sat}\geq1}$, the fraction of the observation time during which $N_{\rm sat}(t)\geq 1$, 
\begin{equation}\label{eq:fnsat}
    f_{N_{\rm sat}\geq1} = \frac{1}{t_{\rm obs}} \int_0^{t_{\rm obs}} \mathbb{1}_{[1, +\infty[} \left( N_{\rm sat}(t) \right) ~ \rm{d}t.
\end{equation}
Here, $f_{N_{\rm sat}\geq1}$ serves as our final statistic in this work. All sampling steps described in this section were repeated a hundred times for each forecast for statistical robustness.

\section{Model validation} \label{app:modelvalidation} 

\begin{figure}[h!]
    \centering
    \includegraphics[width=0.95\linewidth]{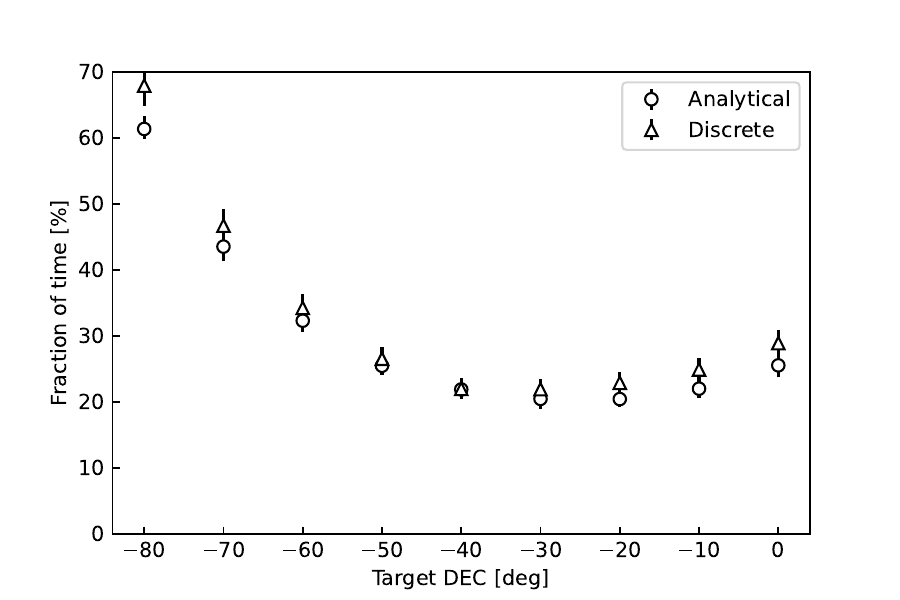}
    \caption{Validation of the analytical model using discrete orbit simulations}
    \label{fig:validation}
\end{figure}
We also validated our analytical modeling against simulations. We started by downloading from \textsc{SpaceTrack}\footnote{\url{www.space-track.org}} the latest available Two-Line Elements (TLEs) for all active Starlink satellite. TLEs contain all orbital parameters necessary to accurately predict orbits within a few days from their acquisition. We assumed 1h observations with SKA-Low, aiming at targets with declinations ranging from 0 to -80$\deg$.  We used $L_{\rm FoV} = 17.8 \degree$, e.g. including the main beam and first sidelobe at 115 MHz. We propagated all orbits and find all satellites that enter the effective beam at least once, and obtained their ingress and egress time. This way, we estimated $f_{N_{\rm sat}\geq1}$. 
\par We ran the analytical model under the same observing assumptions. Regarding the satellite population, we studied the distribution of orbit inclinations obtained from the TLEs. We divided the sample into 4 subsets with $i=43.00, 53.15, 70.00, 97.66\degree$. For the $i=53.15\degree$ subset, we made three distinct shells with altitudes $h=351, 466, 543$ km, while other subsets mainly have constant altitude and hence constitute shells already. This results into the numbers given in Table \ref{tab:shells}. In Figure \ref{fig:validation} we compare the simulation (triangles) and analytical (circles) results. Errorbars display the 1$\sigma$ uncertainty on the estimations. Analytical points lie slightly below the the simulations, but are consistent within 1$\sigma$ of the discrete simulations, except at -80$\degree$ where it is consistent within 2$\sigma$.
\par This test validates the chosen binning of the Starlink sample, that is complex enough to reproduce the results of the simulations. Moreover, we assumed completely random initial ingress time of each satellite into the effective beam: this might infact not be the case as they orbits are coordinated to maximize ground cover. From Figure \ref{fig:validation}, we observe that this assumption does not significantly impact the output of the model compared to TLE simulations. As the continuous coverage targeted by megaconstellations arise from satellites passing anywhere in the night sky, we speculate that the uniform and Poisson sampling are sufficient because the telescope FoV remains small with respect to the half sky.
\par The discrete simulations take 3 hCPU per hour of observations, for 7000 orbits to propagate, while the analytical model take a few sCPU to run. This significant speedup allowed us to compute forecasts for many configurations varying frequency and declination. 

\section{Cumulative effect of all constellations} 
\begin{figure}[h!]
    \centering
    \includegraphics[width=0.95\linewidth]{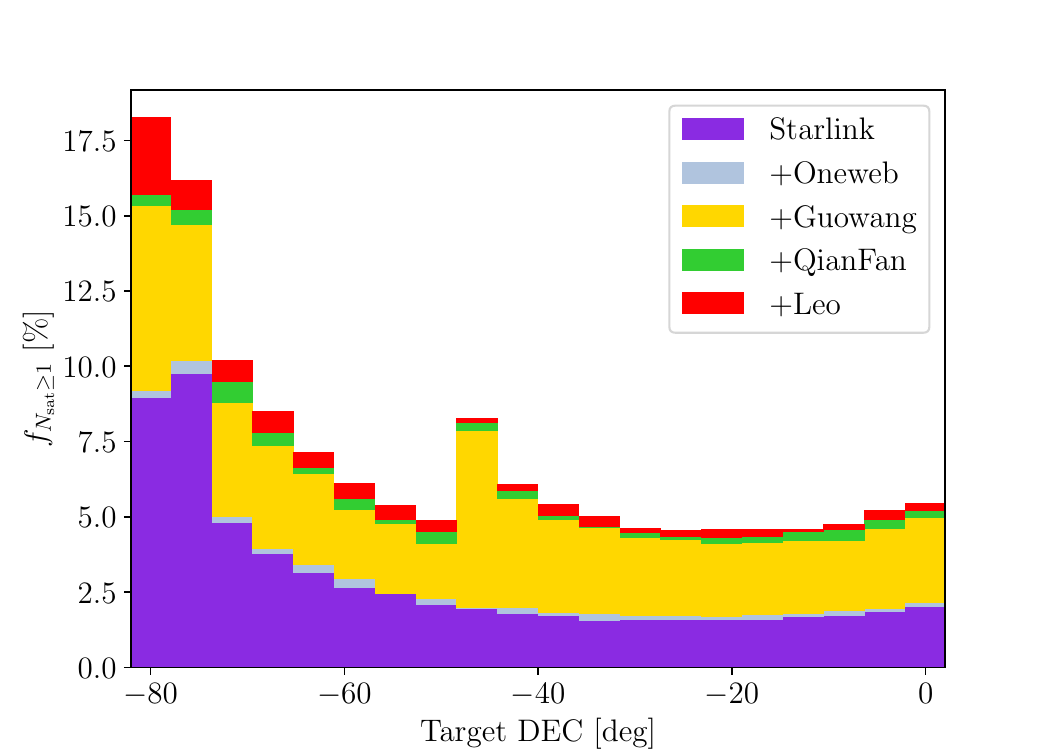}
    \caption{Fraction of time that SKA-Mid will observe at least one satellite in the main beam at 2 GHz. Different colors indicate the successive cumulation of all satellite constellations.}
    \label{fig:cumulative}
\end{figure}
We investigate how different megaconstellations contribute to $f_{N_{\rm sat}\geq1}$. We  take the case of SKA-Mid with $L_{\rm FoV}=2.2 \degree$ (main beam only, at a frequency of 2 GHz). In Figure \ref{fig:cumulative}, we show the cumulative effects of the megaconstellations reported in Table \ref{tab:shells}: purple indicates Starlinks alone, pale blue, yellow and green adds respectively further OneWeb, GuoWang and QianFan, and red completes the picture with Leo. The largest fleets have the biggest impact. However, we note that the altitude also plays an important role, as satellites have lower velocities and therefore spend more time in the beam (see equation \ref{eq:delta_t}). 
\end{appendix}

\end{document}